\documentclass[12pt]{iopart}
\usepackage{epsfig}
\begin{document}
\title{Josephson frequency of resonantly-coupled atomic and molecular condensates}
\author{R A Duine and H T C Stoof}
\address{Institute for Theoretical Physics,
         University of Utrecht, Leuvenlaan 4,\\
         3584 CE Utrecht, The Netherlands}
\ead{\mailto{duine@phys.uu.nl}, \mailto{stoof@phys.uu.nl}}
\begin{abstract}
Motivated by recent experiments [Claussen N R {\it et al.} 2003 {\it
Preprint} cond-mat/0302195] we investigate the magnetic-field dependence
of the Josephson frequency of coherent atom-molecule oscillations near a
Feshbach resonance. Far off resonance this frequency is  determined by
two-body physics. However, the Josephson frequency is relatively large and
a description purely in terms of scattering lengths turns out to be
inadequate. In particular, we have to include the effective range
parameter of the interatomic interaction in our calculations. Close to
resonance, the frequency deviates from the two-body results due to various
many-body shifts. Considering also the many-body effects, we find perfect
agreement with the experimental results over the entire range of magnetic
field.
\end{abstract}
\submitto{\NJP}
\maketitle

\def\bx{{\bf x}}
\def\bk{{\bf k}}
\def\bK{{\bf K}}
\def\half{\frac{1}{2}}
\def\args{(\bx,t)}
\def\argsp{(\bx',t)}
\def\psiup{\hat \psi_{\uparrow}}
\def\psidup{\hat \psi_{\uparrow}^{\dagger}}
\def\psidown{\hat \psi_{\downarrow}}
\def\psiddown{\hat \psi_{\downarrow}^{\dagger}}
\def\psim{\hat \psi_{\rm m}}
\def\psimd{\hat \psi_{\rm m}^{\dagger}}

\section{Introduction}
\label{sec:introduction}
The collisional properties of ultracold alkali atoms are strongly affected
by the presence of a bound state in a coupled closed channel when the energy of
that bound state is close to the energy of the two colliding atoms. This
situation is called a Feshbach resonance \cite{feshbach,stwalley,eite} and
is currently of great experimental interest, because the energy difference
between the two colliding atoms and the molecular bound state is 
magnetic-field dependent due to the different Zeeman shifts of the atoms
in the open channel and the molecule. This leads to a high level of
experimental control over the interatomic interactions \cite{inouye}.

The physics of a Feshbach resonance bears some similarity with an ordinary
quantum-mechanical two-level system, where one level represents the
two atoms and the other level the molecular state. Inspired by this
picture Donley {\it et al.} have conducted a Ramsey-type experiment, by
starting from an essentially pure atomic condensate and applying two short
pulses in the magnetic field separated by a longer evolution time
\cite{JILA1}. As a function of the evolution time oscillations in the
number of remaining condensate atoms are observed. Sufficiently far from
resonance the frequency corresponds to the molecular binding energy, in
agreement with the fact that in that case the coupling may be neglected
and the Josephson frequency of the oscillations between an atomic and a
molecular Bose-Einstein condensate is equal to the energy difference
between the two atoms and the molecule. More recent high-precision
experiments by Claussen {\it et al.}, over a larger range of magnetic
fields, have shown that close to the resonance the frequency deviates from
the two-body result as a result of many-body effects \cite{claussen2003}. 

In order to be able to theoretically describe the magnetic-field
dependence of the observed frequency of coherent atom-molecule
oscillations over a large range of magnetic field, we have to incorporate
into the many-body theory the relevant two-atom physics exactly, and, in
particular therefore, the correct molecular binding energy. Far off
resonance this binding energy is relatively large and it turns out to be
necessary to include the energy dependence of the interactions and
coupling constants in the theoretical description. In particular, the use
of solely the scattering length is inadequate and we also need to include
the effective range of the interatomic interaction potential. In addition,
to describe the deviation from the two-body result close to resonance, the
theory has to incorporate the correct mean-field interactions between
condensate and noncondensate atoms that ultimately give rise to mean-field
shifts of the molecule. Although a large number of important contributions
have been made in formulating a theory for Feshbach-resonant interactions
\cite{peter,eddy,murray1,servaas,juha,keith,rembert4}, it appears that
none of the existing theories provides a quantitative agreement with the
experimentally observed Josephson frequency over the full range of the
magnetic field. It is the aim of this paper to put forward such a theory
and thereby also improve upon our previous calculations which were
restricted to the region close to resonance \cite{rembert6}. 

To this end, the paper is organized as follows. In \sref{sec:atommolecule}
we first present the many-body theory that incorporates the molecular
binding energy exactly. This theory includes the effective range of the
interatomic interactions, which is determined from experiment in
\sref{sec:binding} by calculating the molecular binding energy in vacuum.
In \sref{sec:josephson} we calculate the frequency of the coherent
atom-molecule oscillations, i.e., the Josephson frequency that includes
the many-body shifts of the two-body results. We end in
\sref{sec:conclusions} with our conclusions. 

\section{Atom-molecule coherence}
\label{sec:atommolecule}
In this section we present the theory for the description of
Feshbach-resonant interactions between atoms. The theory presented here
improves upon our results presented in \cite{rembert4} by including the
energy dependence of the various coupling constants and, in particular
therefore, the effective range of the atom-atom interactions. 
\subsection{Microscopic theory}
\label{subsec:bare}
Our starting point is the microscopic atom-molecule hamiltonian for the
description of a Feshbach resonance \cite{rembert4}. For a homogeneous
system in a box of volume $V$ this hamiltonian reads 
\begin{eqnarray}
\label{eq:bareatommol}
  \hat H = \sum_{\bk} \epsilon_\bk \hat a^{\dagger}_\bk \hat a_\bk
    + \frac{1}{2V} \sum_{\bK,\bk,\bk'}V (\bk-\bk') \hat a^{\dagger}_{\bK/2+\bk} \hat
    a^{\dagger}_{\bK/2-\bk} \hat a_{\bK/2+\bk'} \hat
    a_{\bK/2-\bk'} \nonumber \\
    + \sum_{\bk} \left[ \frac{\epsilon_\bk}{2} + \delta_{\rm B} (B) \right] \hat b^{\dagger}_\bk \hat b_\bk
    + \frac{1}{\sqrt{V}} \sum_{\bK,\bk} g_{\rm B} (\bk) \left[ \hat b^{\dagger}_\bK \hat
    a_{\bK/2+\bk} \hat a_{\bK/2-\bk} + {\rm h.c.} \right]~.
\end{eqnarray}
Here, the operator $\hat a^{\dagger}_\bk$ creates an atom with momentum
$\hbar \bk$, and its hermitian conjugate (h.c.) annihilates an atom with
this momentum. The single-atom dispersion is given by $\epsilon_\bk =
\hbar^2 \bk^2/2m$. The molecules are described by the operators $\hat
b^{\dagger}_\bk$ and $\hat b_\bk$ and have a single-particle dispersion
given by $\epsilon_\bk/2+\delta_{\rm B} (B)$ since their mass is twice the
atomic mass $m$. Here $\delta_{\rm B} (B)$ is the detuning of the bare
molecules, i.e., the energy difference between the bare molecules and the
atoms, which depends on the magnetic field $B$ due to the fact that the
magnetic moment of the molecule is different from that of the atoms.  The
Fourier transform of the atom-atom interaction in the incoming channel is
given by $V (\bk)$, and goes to zero as $\bk \to \infty$ due to the
nonzero range of the atomic interactions. The bare  atom-molecule coupling
is given by $g_{\rm B} (\bk)$, and also goes to zero as $\bk \to \infty$,
due to the fact that the wave function of the bare molecule has a finite,
nonzero extent. Both the atom-molecule interactions, as well as the
molecule-molecule interactions have been neglected, because we are
interested in the Josephson oscillations of an almost pure atomic
condensate.

In applying the above microscopic hamiltonian to realistic atomic gases we
have to perform perturbation theory in the atom-atom interaction and the
atom-molecule coupling. These quantities are large and we have to consider
them to all orders. At the low densities of interest to us here, we need
only to consider all ladder diagrams, since three and more body processes
are negligible. This automatically builds into our theory the relevant
two-atom physics, i.e., the scattering amplitude of the atoms and the
binding energy of the molecule. Note that the fact that both the
atom-molecule coupling and the atom-atom interaction decrease rapidly with
increasing momenta ensures that no ultraviolet divergencies are
encountered when doing perturbation theory in these quantities. In the
next section we discuss the required  perturbative expansion.
\subsection{Ladder diagrams}
\label{subsec:ladders}
\begin{figure}
\begin{center}
\epsfbox{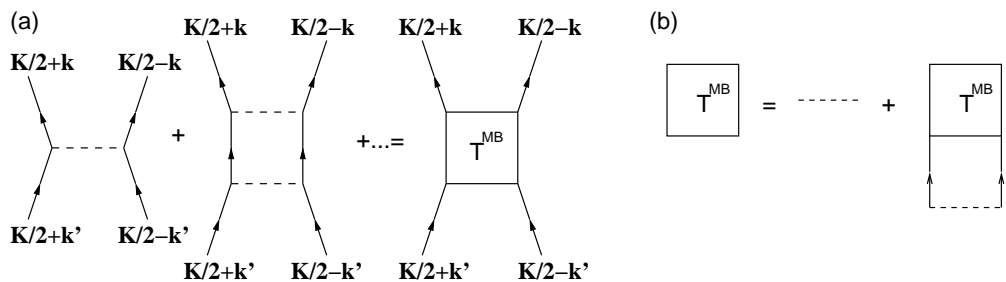}
\end{center}
\caption{\label{fig:tmatrix} (a) Born series for the scattering amplitude.
(b) Diagrammatic representation of the
many-body T matrix. The solid lines correspond to single-atom
propagators. The dashed lines corresponds to the interatomic
interaction $V(\bk)$.}
\end{figure}

First, we consider the atom-atom scattering amplitude. To lowest order in
the interaction the amplitude of two atoms scattering from relative
momentum $\hbar \bk'$ to relative momentum $\hbar \bk$ is simply equal to
$V(\bk-\bk')$. However, since this interaction is strong we have to
consider also the situation where the atoms collide twice, three times and
so on. Diagrammatically, this leads to the Born series shown in
\fref{fig:tmatrix}~(a), which is summed by introducing the many-body
T(ransition) matrix that obeys the Bethe-Salpeter equation
\begin{eqnarray}
\label{eq:mbtmatrix}
  T^{\rm MB} (\bk,\bk',{\bf K},z) &=& V (\bk-\bk')
   \nonumber \\  &+& \frac{1}{V} \sum_{\bk''} V (\bk-\bk'')
    \frac{ \left[ 1+ N(\epsilon_{{\bf K}/2+\bk''}-\mu)
                  + N(\epsilon_{{\bf K}/2-\bk''}-\mu)\right]}
         {z-\epsilon_{{\bf K}/2+\bk''}-\epsilon_{{\bf K}/2-\bk''}}
     \nonumber \\  && \times T^{\rm MB} (\bk'',\bk',{\bf K},z)~.
\end{eqnarray}
Here we have used the Hartree-Fock approximation for the thermal atoms and, for the
moment, neglected mean-field shifts. Furthermore, $N(x)=[e^{\beta x}-1]^{-1}$ is the
Bose distribution function of the atoms, $\mu$ their chemical potential,
and $\beta=1/k_{ \rm B}T$ the inverse thermal energy. The total energy at which the
scattering takes place is denoted by $z$. Due to the surrounding medium the
scattering amplitude depends also on the center-of-mass momentum $\bK$ of
the colliding atoms. The diagrammatic representation of this equation is
given in \fref{fig:tmatrix}~(b).

For temperatures not too close to the critical temperature we are allowed
to neglect the many-body effects \cite{stoof1}, and \eref{eq:mbtmatrix}
reduces to the Lippmann-Schwinger equation for the two-body T matrix. The
effective scattering amplitude thus becomes $T^{\rm 2B}({\bf
k},{\bf k}',z-\epsilon_{\bf K}/2)$. For the small external momenta of
interest to us here the momentum dependence of the two-body T matrix may be
neglected. Its energy dependence, however, cannot be neglected, since we
are ultimately interested in calculating the molecular binding energy
which is relatively large far off resonance. Including the energy dependence of
the T matrix \cite{stoof1988}, we thus conclude that the renormalization
of the amplitude for two atoms scattering at the physically relevant
energy $\hbar \omega^+\equiv\hbar \omega+i0$ is given by
\begin{equation}
\label{eq:vren}
  V (\bk-\bk') \to T^{\rm 2B} ({\bf 0},{\bf 0}, \hbar \omega^+)
  =\frac{4 \pi a_{\rm bg} \hbar^2}{m} 
    \left[ \frac{1}{1+ia_{\rm bg} \sqrt{\frac{m \omega}{\hbar}}-\frac{a_{\rm bg}
    r_{\rm bg} m \omega}{2 \hbar}}\right]~.
\end{equation}
Here $a_{\rm bg}$ denotes the $s$-wave scattering length of the potential
$V(\bk)$, which, due to the fact that it describes the nonresonant part of
the atomic interactions, is called the background scattering length. The
parameter $r_{\rm bg}$ is, by definition, the effective range of this potential.

\begin{figure}
\begin{center}
\epsfbox{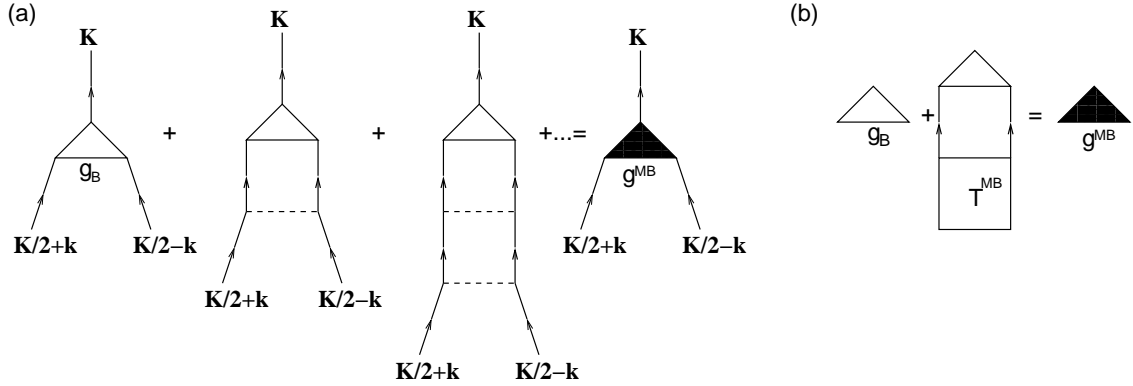}
\end{center}
\caption{\label{fig:gren} Diagrammatic calculation of the amplitude of two
atoms to form a molecule. The open and filled triangles represent the bare
and many-body coupling constant between the atoms and the molecule,
respectively.}
\end{figure}

Next, we calculate the amplitude for two atoms with momenta $\hbar
\bK/2+\hbar \bk$ and $\hbar \bK/2 -\hbar \bk$ to form a molecule with energy $z$. To
lowest order in the atomic interaction this amplitude is equal to $g_{\rm
B} (\bk)$. However, before forming a molecule, the two atoms may undergo
multiple collisions and we again have to take into account the full Born
series, diagrammatically given in \fref{fig:gren}~(a), which is summed by
again introducing the many-body T matrix as shown in \fref{fig:gren}~(b).
This leads to the equation 
\begin{eqnarray}
\label{eq:rengb}
  g^{\rm MB} (\bk,{\bf K},z) &=& g_{\rm B}(\bk)
  + \frac{1}{V} \sum_{\bk'}
   T^{\rm MB} (\bk,\bk',{\bf K},z) \nonumber \\
   &&\times \frac{\left[ 1+ N(\epsilon_{{\bf K}/2+\bk'}-\mu)
                          + N(\epsilon_{{\bf K}/2-\bk'}-\mu)\right]}
                 {z-\epsilon_{{\bf K}/2+\bk'}-\epsilon_{{\bf K}/2-\bk'}}
    g_{\rm B}(\bk')~.
\end{eqnarray} 
Neglecting many-body effects, the amplitude becomes $g^{\rm
2B}(\bk,z-\epsilon_{\bf K}/2)$ with
\begin{eqnarray}
\label{eq:rengb2b}
  g^{\rm 2B} (\bk,z) = g_{\rm B}(\bk) + \frac{1}{V} \sum_{\bk'}
   T^{\rm 2B} (\bk,\bk',z) \frac{1}{z-2\epsilon_{\bf k'}}
    g_{\rm B}(\bk')~.
\end{eqnarray}
From the above equation we infer that the energy dependence of the
amplitude is the same as that of the two-body T matrix. This result is 
easily understood by noting that for a contact potential $V(\bk)=V_{\bf
0}$  and we simply have that $g^{\rm 2B} = g_{\rm B} T^{\rm 2B}/V_{\bf 0}$. Hence
we have for the renormalization of the amplitude for forming a molecule
with energy $\hbar \omega$ the expression
\begin{equation}
\label{eq:gedep}
   g_{\rm B} \to g^{\rm 2B} ({\bf 0},\hbar \omega^+) =g
   \left[ \frac{1}{1+ia_{\rm bg} \sqrt{\frac{m \omega}{\hbar}}-\frac{a_{\rm bg}
    r_{\rm bg} m \omega}{2 \hbar}}\right]~,
\end{equation}
where $g$ is the atom-molecule coupling constant at zero energy. 

Finally, we have to take into account the fact that the coupling between the atoms
and the molecule gives the molecule a finite lifetime and leads to a change
in the bound-state energy of the molecule. Physically, this is described by
the self-energy of the molecules which, within our approximations, is equal
to $\hbar \Sigma^{\rm 2B}_{\rm m} (z-\epsilon_\bK/2)$ with
\begin{equation}
\label{eq:self}
  \hbar \Sigma_{\rm m}^{\rm 2B} (z) =2 \int \frac{d\bk}{(2 \pi)^3} 
  \left|g^{\rm 2B}({\bf 0},2 \epsilon_\bk)\right|^2~ \frac{1}{z- 2
  \epsilon_\bk}.
\end{equation} 
This expression is easiest understood by considering the fact that
evaluated at positive energy $\hbar \omega^+$ and using ${\rm Im}
[1/(\omega^{\pm}-\omega')]=\mp \pi \delta (\omega-\omega')$, it 
immediately leads to the correct Fermi's Golden Rule for the decay rate of the
molecule. Using the result for the amplitude in \eref{eq:gedep} the
retarded self-energy $\hbar \Sigma^{(+)}_{\rm m} (\hbar \omega)$, i.e.,
the selfenergy evaluated at energy $\hbar \omega^+$, becomes
\begin{equation}
\label{eq:selfmret}
 \hbar \Sigma^{(+)}_{\rm m} (\hbar \omega) 
   =-\frac{g^2 m}{2 \pi \hbar^2 \sqrt{1-2\frac{r_{\rm bg}}{a_{\rm bg}}}} 
   \left[ \frac{i \sqrt{\left(1-2\frac{r_{\rm bg}}{a_{\rm bg}}\right) 
   \frac{m \omega}{\hbar}}-
    \frac{r_{\rm bg} m \omega}{2 \hbar}}
   {1+ia_{\rm bg}\sqrt{\left(1-2\frac{r_{\rm bg}}{a_{\rm bg}}\right)\frac{m\omega}{\hbar}}
   - \frac{r_{\rm bg} a_{\rm bg} m \omega}{2 \hbar}}
   \right]~,
\end{equation}
where we have for simplicity assumed that the background scattering length
is negative, which is the case for the applications of interest here. We
have also subtracted the $z=0$ part in the integral in \eref{eq:self}
since this shift renormalizes the bare detuning according to $\delta_{\rm
B} (B) \to \delta (B)$. Note that the square-root behaviour of $\hbar
\Sigma^{(+)}_{\rm m} (\hbar \omega)$ at small energies is in agreement
with the expected Wigner-threshold law for the molecule to decay into the
two-atom continuum.

At this point we make the connection with experimentally known parameters.
The resonance is characterized experimentally by a width $\Delta B$ and a
position $B_0$. More precisely, the $s$-wave scattering length of the
atoms as a function of magnetic field is given by 
\begin{equation}
\label{eq:ascatofb} 
  a(B) = a_{\rm bg} \left( 1-\frac{\Delta B}{B-B_0} \right). 
\end{equation} 
Therefore, in order to reproduce the experimentally observed width of the
resonance we have that $g=\hbar \sqrt{2 \pi a_{\rm bg} \Delta B  \Delta
\mu /m}$, since an elimination of the molecular field shows that $a_{bg}
\Delta B = m g^2/(2 \pi \hbar^2 \Delta \mu)$. We have made use of the fact
that the detuning is given by $\delta (B)=\Delta \mu (B-B_0)$ where
$\Delta \mu$ is the difference in magnetic moment between two atoms and a
bare molecule.

In this paper we focus on the experiments by Claussen {\it et al.}
\cite{claussen2003}. In these experiments the Feshbach resonance
at $B_0=155.041(18)$ G(auss) in the $|f=2;m_f=-2\rangle$ state of
$^{85}$Rb is used. The width of the resonance is equal to $\Delta=10.71(2)$ G and the
background scattering length is $a_{\rm bg}=-443 a_0$ with $a_0$ the Bohr
radius. The difference in magnetic moment between two atoms and the bare
molecule is equal to $\Delta \mu =-2.23 \mu_{\rm B}$ with $\mu_{\rm B}$
the Bohr magneton \cite{servaas}. In the next section we also determine the
effective range of the atom-atom interactions by calculating the binding
energy of the molecule and comparing the result with experiment.

\section{Molecular binding energy}
\label{sec:binding}
The molecular bound-state energy is determined by solving for $\hbar
\omega$ in the equation
\begin{equation}
\label{eq:poles}
  \hbar \omega = \delta (B) + \hbar \Sigma_{\rm m}^{(+)}
  (\hbar \omega)~.
\end{equation} 
For positive detuning this leads to an imaginary bound-state energy, in
agreement with the fact that the molecule decays when its energy is
above the threshold of the two-atom continuum. For negative detuning there is a
bound state with energy $\epsilon_{\rm m} (B)$. Close to resonance we are
allowed to neglect the contribution from the effective range to the
self-energy in \eref{eq:selfmret} and we solve \eref{eq:poles}
analytically to find 
\begin{equation}
\label{eq:ebres}
  \epsilon_{\rm m}(B)  \simeq - \frac{\hbar^2}{m [a(B)]^2}~.
\end{equation}
This is indeed the correct result for the molecular bound state energy
close to resonance \cite{servaas}. 

\begin{figure}
\begin{center}
\epsfbox{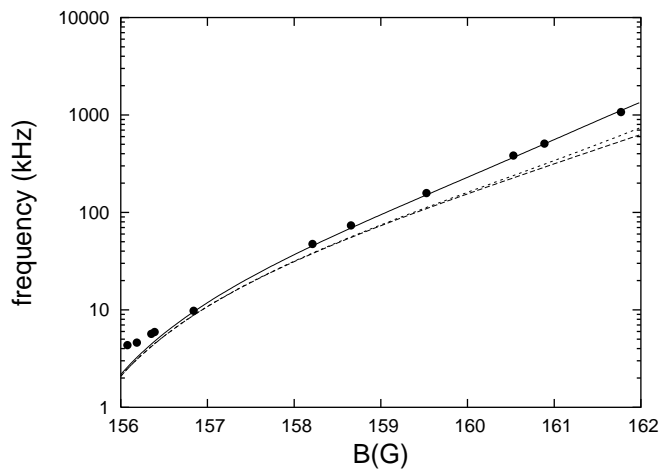}
\end{center}
\caption{\label{fig:twobody} Molecular binding energy in vacuum. The solid
line shows the result of a calculation with $r_{\rm bg}=185 a_0$, the
dashed line shows the result for $r_{\rm bg}=0$. The dotted line shows 
$|\epsilon(B)|=\hbar^2/ma^2$. The experimental points are taken
from \cite{claussen2003}.}
\end{figure}

Far off resonance, we are no longer allowed to neglect the effective range
of the interatomic interaction. Because of this, the equation for the
bound-state energy can not be solved analytically anymore, but is
nevertheless easily solved numerically. In order to determine the
effective range we compare the result with the experimental results of
Claussen {\it et al.} \cite{claussen2003}. In these experiments  the
Josephson oscillation frequency for coherent atom-molecule oscillations is
determined as a function of the magnetic field. Far off resonance this
frequency is essentially equal to the molecular binding energy and
therefore independent of the condensate density. By comparing our results
with the experimental data far off resonance, we determine the effective
range uniquely. In \fref{fig:twobody} we show the binding energy as a
function of the magnetic field. The solid line is the result of a
calculation with $r_{\rm bg} = 185 a_0$. The dashed line shows the result
for zero effective range and the dotted line shows the magnetic-field
dependence given in \eref{eq:ebres}. The experimental points, taken from
\cite{claussen2003}, are also shown. It should be noted that the errorbars
on the experimental data are roughly indicated by the size of the points.
Clearly, the result with zero effective range, as well as the expression
given in \eref{eq:ebres} deviate significantly from the experimental data
for magnetic fields larger than $158$ G. Interestingly, the result for
zero effective range deviates more from the experimental result than the
expression given by \eref{eq:ebres}. The curve obtained with an
effective range of $r_{\rm bg}=185 a_0$ agrees very well with the
experimental data in the regime where $B>157$ G and hence we use this
value for the effective range of the interatomic interaction from now on.
As expected, close to the resonance, the results of the three calculations
become identical because the energy dependence of the effective
interactions becomes less important in this regime. Note, however, that
all three results deviate from the experimental points for magnetic fields
smaller than $157$ G. This deviation is explained by considering many-body
effects which are the topic of the next section.
 
\section{Josephson oscillations}
\label{sec:josephson}
\begin{figure}
\begin{center}
\epsfbox{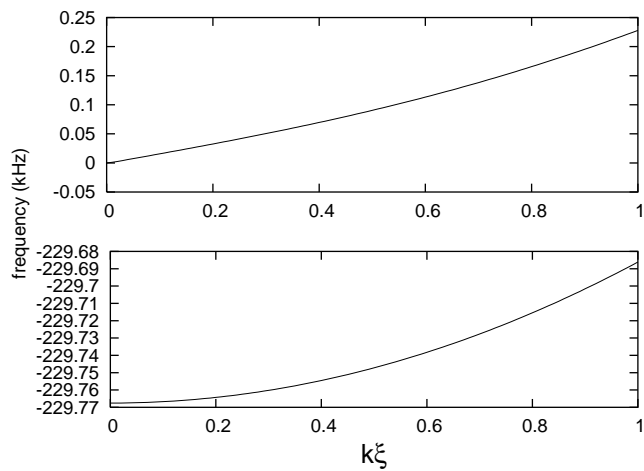}
\end{center}
\caption{\label{fig:disp} Dispersion relation of the collective modes of
an atom-molecule system for a condensate density of $n_{\rm
a}=2\times10^{12}$ cm$^{-3}$ at a magnetic field of $B=160$ G. The upper
branch corresponds to the gapless dispersion for phonons. The lower
branch corresponds to coherent atom-molecule oscillations. The momentum
is indicated in units of the inverse coherence length $\xi^{-1}=\sqrt{16 \pi a n_{\rm
a}}$.}
\end{figure}
To determine the Josephson frequency of coherent oscillations between the
atomic and the molecular condensate we calculate the excitation spectrum of
the system by means of linear response theory. Therefore, we write for the
atomic and molecular Heisenberg creation operators
\begin{eqnarray}
\label{eq:linearization}
  \hat a_\bk (t) \!&=&\!\left[ \sqrt{N_{\rm a}}
    +u_{\bf k} e^{-i \omega t} \hat \alpha_\bk
    -{v_{-\bf k}} e^{+i \omega t} \hat \alpha^{\dagger}_\bk \right]e^{-i \mu
 t/\hbar}~, {\rm and}
 \nonumber \\
  \hat b_\bk (t) \!&=&\!\left[ \sqrt{N_{\rm m}} +u'_{\bf k}
e^{-i \omega t} \hat \beta_\bk
 - v'_{-\bf k} e^{+i \omega t} \hat \beta^{\dagger}_\bk \right]e^{-i 2 \mu
 t/\hbar}~,
\end{eqnarray}
respectively. The operators $\hat \alpha_\bk$ and $\hat \beta_\bk$
describe a Bogoliubov quasi-atom and quasi-molecule, respectively, and in
linear response theory we keep terms up to quadratic order in these
operators. The number of particles in the atomic and molecular condensate
is equal to $N_{\rm a}$ and $N_{\rm m}$, respectively. In the presence of
an atomic condensate it is crucial to take into account its mean-field
effects on the thermal atoms. This mean-field energy is, in the
Hartree-Fock approximation, given by
\begin{eqnarray} 
\label{eq:sigmahf}
\hbar \Sigma^{\rm HF} =
 2 n_{\rm a} \left( \rule{0mm}{9mm} T^{\rm 2B} \left({\bf 0},{\bf 0},\mu -\hbar \Sigma^{\rm
 HF}\right) \right. \nonumber \\ \left.
 +\frac{2 \left|g^{\rm 2B}\left({\bf 0},{\bf 0},\mu-\hbar \Sigma^{\rm
  HF}\right)\right|^2}{\hbar \Sigma^{\rm HF}+\mu
 -\delta (B) - \hbar \Sigma_{\rm m}^{(+)}\left(\mu-\hbar \Sigma^{\rm HF} \right) 
 } \right)~,
\end{eqnarray}  
and reduces to the usual expression $8 \pi a(B) \hbar^2 n_{\rm a}/m$ far off
resonance. Here, $n_{\rm a} \equiv N_{\rm a}/V$ denotes the density of the
atomic Bose-Einstein condensate.

To find the excitation spectrum, we have to diagonalize the hamiltonian in
terms of the Bogoliubov quasi-particle operators. This requires the
diagonalization of a $4\times4$ matrix equation for the coherence factors
$u_\bk, v_\bk, u'_\bk$ and $v'_\bk$. However, the coherence factors for the
molecular operators are straightforwardly eliminated. This yields the
$2\times2$ eigenvalue problem given by
\begin{eqnarray}
\label{eq:eigensystem}
  \left[T^{2B}_{\rm eff}(2\mu) n_{\rm a} \right]^* u_\bk
    +\left[  \epsilon_\bk - \mu + 2T^{\rm 2B}_{\rm eff} 
      (2\mu-\hbar\omega-\epsilon_\bk/2) n_{\rm a}  \right]^*
      v_\bk=-\hbar \omega v_\bk \nonumber \\
   \left[ \epsilon_\bk - \mu + 2 T^{\rm 2B}_{\rm eff} 
      (2\mu+\hbar\omega-\epsilon_\bk/2) n_{\rm a} \right] u_\bk
      +T^{\rm 2B}_{\rm eff} (2 \mu)n_{\rm a} v_\bk = \hbar \omega
      u_\bk~,
\end{eqnarray} 
where the effective atom-atom interaction is given by
\begin{eqnarray}
\label{eq:t2beff}
  T^{\rm 2B}_{\rm eff} (\hbar \omega) = 
  T^{\rm 2B} \left({\bf 0},{\bf 0},\hbar \omega^+ -2 \hbar \Sigma^{\rm
  HF}\right)
  \nonumber \\  \ \ \ \ \ \ \ \ \ \ 
  +\frac{2\left|g^{\rm 2B}\left({\bf 0},{\bf 0},\hbar \omega^+-2 \hbar \Sigma^{\rm
  HF}\right)\right|^2}
  {\hbar \omega^+ - \delta (B) - \hbar
  \Sigma^{(+)}_{\rm m} \left(\hbar \omega - 2 \hbar \Sigma^{\rm HF}\right)}~.
\end{eqnarray}
The chemical potential is determined by the time-independent
Gross-Pitaevskii equation, which, for energy-dependent interactions, reads
\begin{equation}
\label{eq:tidptgpe}
  \mu = T^{\rm 2B}_{\rm eff} (2\mu) n_{\rm a}~.
\end{equation}
Note that this last equation and \eref{eq:sigmahf} uniquely
determine the chemical potential and the Hartree-Fock mean-field energy
for a given atomic condensate density. Also note that it is crucial to
include the Hartree-Fock mean-field shift to have equilibrium solutions.
The number of molecules in the molecular condensate can be calculated from
the number of condensate atoms, but is eliminated here to get the effective
atom-atom interaction \cite{rembert6}.

\begin{figure}
\begin{center}
\epsfbox{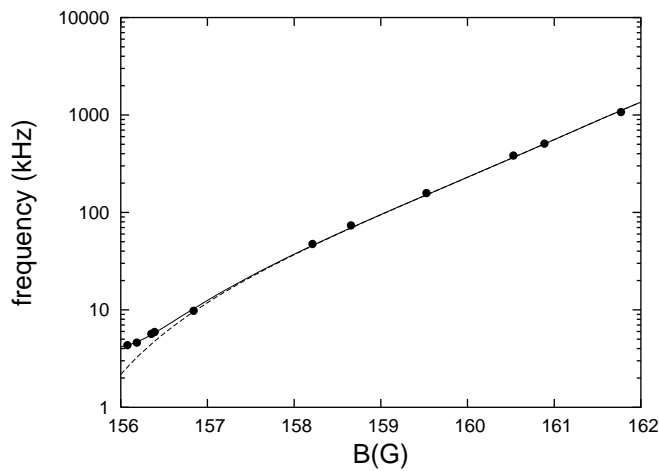}
\end{center}
\caption{\label{fig:joseph} Josephson frequency of coherent atom-molecule
oscillations for an atomic condensate density of $n_{\rm a}=2 \times
10^{12}$ cm$^{-3}$ . The solid line shows the result of the calculation of the
Josephson frequency.
The dashed line shows the molecular binding energy in vacuum. 
The experimental points are taken from \cite{claussen2003}.}
\end{figure}

The dispersion relation of the collective excitations is found by solving
the eigenvalue problem in \eref{eq:eigensystem} numerically. This yields
two branches, one gapless branch corresponding to the phonon modes, and a
gapped branch corresponding to the atom-molecule oscillations. Physically,
the difference between the two branches is understood by realizing that
for the phonon modes the phases of the atomic and the molecular condensate
are locked to each other and oscillate in phase. Since the hamiltonian is
invariant under transformations $\hat a \to \hat a e^{i \theta}$ and $\hat
b \to \hat  b e^{2i \theta}$ we conclude that the phonons are indeed
gapless, and, in fact, correspond to the
Goldstone mode associated with the breaking of the $U(1)$ symmetry by the
condensates. For the coherent atom-molecule oscillations the phases of the
atomic and molecular condensate oscillate out of phase and hence the
associated dispersion is gapped. \Fref{fig:disp} shows the two branches of
the dispersion for a condensate density of $n_a=2 \times 10^{12}$ cm$^{-3}$ at
a magnetic field of $B=160$ G.

The zero-momentum part of the gapped branch corresponds to the Josephson
frequency of the coherent oscillations between the atomic and the
molecular condensate, as observed by Claussen {\it et al.}
\cite{claussen2003}. In \fref{fig:joseph} the dotted line shows the result
of the calculation of this frequency for a condensate density of $n_a=2
\times 10^{12}$ cm$^{-3}$. This density corresponds to the effective
homogeneous density we have to take in order to compare to the
experimental results found by Claussen {\it et al.}, which are done in an
inhomogeneous magnetic trap \cite{claussen2003}. Roughly speaking, this
comes about because in the comparison of our homogeneous calculation with
the inhomogeneous situation the effective Josephson coupling, which mostly
determines the deviations from the two-body result, is reduced by an
overlap integral of the atomic and molecular wave functions. This reduction
factor multiplied by the experimental central density determines the
effective homogeneous density \cite{rembert6}. Clearly, our result shows
perfect agreement with the experimental data points over the entire
experimentally investigated range of the magnetic field. The dashed line
in \fref{fig:joseph} corresponds to the molecular binding energy in
vacuum, also shown in \fref{fig:twobody}. As expected, the Josephson
frequency becomes equal to the binding energy far off resonance.

\section{Conclusions}
\label{sec:conclusions}
We have improved upon our previous work \cite{rembert4,rembert6} in
calculating the Josephson frequency of coherent atom-molecule
oscillations. In doing so, we had to include the energy-dependence of the
interactions, and, in particular, the effective range of the interatomic
interactions. We have found excellent agreement with the available
experimental results. In future work we intend to study the density and
magnetic-field dependence of the Josephson frequency in more detail.

\ack It is a pleasure to thank Neil Claussen for providing us with the
experimental data obtained by the group of Carl Wieman.


\section*{References}
\begin{thebibliography}{99}
\bibitem{feshbach} Feshbach H 1962 {\it Ann. Phys.} {\bf 19} 287
\bibitem{stwalley} Stwalley W C 1976 {\it Phys. Rev. Lett.} {\bf 37} 1628
\bibitem{eite} Tiesinga E, Verhaar B J and Stoof H T C 1993
               {\it Phys. Rev. A} {\bf 47} 4114
\bibitem{inouye} Inouye S, Andrews M R, Stenger J, Miesner H J,
Stamper-Kurn D M, Ketterle W 1998 {\it Nature} {\bf 392} 151	      
\bibitem{JILA1} Donley E A, Claussen N R, Thompson S T and Wieman C E
                2002 {\it Nature} {\bf 417} 529
\bibitem{claussen2003} Claussen N R, Kokkelmans S J J M F, Thompson S T
Donley E A, Wieman C E 2003 {\it Preprint} cond-mat/0302195
\bibitem{peter} Drummond P D, Kheruntsyan K V and He H 1998
                {\it Phys. Rev. Lett.} {\bf 81} 3055
\bibitem{eddy} Timmermans E, Tommasini P, C\^ot\'e R, Hussein M and
               Kerman A 1999 {\it Phys. Rev. Lett.} {\bf 83} 2691
\bibitem{murray1} Holland M, Park J and Walser R
                  2001 {\it Phys. Rev. Lett.} {\bf 86} 1915
\bibitem{servaas} Kokkelmans S J J M F and Holland M
                  2002 {\it Phys. Rev. Lett.} {\bf 89} 180401
\bibitem{juha} Mackie M, Suominen K A and Javanainen J
               2002  {\it Phys. Rev. Lett.} {\bf 89} 180403
\bibitem{keith} K\"ohler T, Gasenzer T and Burnett K
                2002 {\it Phys. Rev. A} {\bf 67} 013601	
\bibitem{rembert4} Duine R A and Stoof H T C. 2003 {\it Preprint}
cond-mat/0210544, to be published in {\it J. Opt. B: Quantum Semiclass.
Opt.} {\bf 5}.	
\bibitem{rembert6} Duine R A and Stoof H T C 2003 {\it Preprint}
cond-mat/0302304
\bibitem{stoof1} Stoof H T C, Bijlsma M and Houbiers M 1996
                 {\it J. Res. Natl. Inst. Stand. Technol.} {\bf 101} 443
\bibitem{stoof1988} {\it See, for instance,} 
Stoof H T C, de Goey L P H, Rovers W M H M, Kop Jansen
P S M, Verhaar B J 1988 {\it Phys. Rev. A.} {\bf 38} 1248		
		
		



\end{thebibliography}
\end{document}